\documentclass[a4paper,11pt,twoside]{article}
\usepackage{graphicx,ep1-mydefs}
\begin{document}

\title{\vspace{-1cm}\hfill \small KUCP0154\\[1cm]
\sc\huge 
Pareto's Law for Income of Individuals and Debt of Bankrupt Companies}
\author{
{\Large\sc Hideaki Aoyama$^\dagger$, 
Yuichi Nagahara$^*$, 
}\\[6pt]
{\Large\sc 
Mitsuhiro P. Okazaki$^\sharp$,
Wataru Souma$^\dagger$,
}\\[6pt]
{\Large \sc 
Hideki Takayasu$^\ddagger$,
{\rm and} 
Misako Takayasu$^\star$ 
}\\[8pt]
$^\dagger$Faculty of Integrated Human Studies\\[3pt]
Kyoto University, Kyoto 606-8501, Japan\\[6pt]
$^*$School of Political Science and Economics, Meiji University,\\[3pt]
1-1 Kanda-Surugadai, Chiyoda-ku, Tokyo 101-8301, Japan\\[6pt]
$^\sharp$Factory Automation System Department, 
Takagi-Shokai Co.,\\[3pt]
2-2-7 Kitasennzoku, Ohta-ku, Tokyo 145-0062, Japan\\[6pt]
$^\ddagger$Sony Computer Science Laboratories,\\[3pt]
3-14-13 Higashi-Gotanda, Shinagawa-ku, Tokyo 141-0022, Japan\\[6pt]
$^\star$Department of Complex Systems, Future University -- Hakodate,\\[3pt]
116-2 Kameda-Nakano-cho, Hakodate, Hokkaido 041-8655, Japan.}

\date{May 25, 2000}
\maketitle
\thispagestyle{empty}
\begin{abstract}
We analyze the distribution of income and income tax of 
individuals in Japan for the fiscal year 1998.
From the rank-size plots we find that the accumulated probability distribution
of both data obey a power law with a Pareto exponent 
very close to $-2$.
We also present an analysis of the distribution of the debts owed by
bankrupt companies from 1997 to March, 2000,
which is consistent with 
a power law behavior with a Pareto exponent equal to $-1$.
This power law is the same as that of the income distribution of companies.
Possible implications of these findings for model building are discussed.
\end{abstract}

\newpage
\section{Introduction}
More than a century ago the Italian sociologist Pareto studied the
distribution of personal incomes for the purpose of characterizing a whole
country's economic status \cite{pareto}. He found power law cumulative
distributions with exponents close to $-1.5$ for several countries, which
turned out to be a classic example of fractal distributions \cite{second}. 
In 1922 Gini checked the same statistics and reported that power laws actually
hold, but the values of the exponents vary from country to country
 \cite{gini}. 
Montroll and
Shlesinger analyzed the USA's personal income data for the year 1935--36 
and found
that the top 1\% of incomes follow a power law with an exponent $-1.63$,
while
the rest, who are expected to be salaried, follow a log-normal
distribution \cite{lognormal}. 
Although these results are interesting and
suggestive, the data are all old, and a much more
precise analysis using contemporary high quality data in digital form
is desirable. One
of this paper's aims is to answer this requirement by showing the results of
personal income distribution analysis on the 
latest high precision data reported
to the tax office in Japan. As we will show in section 2 we find a clear
power law with an exponent very close to $-2$, the first time
it has been reported in this kind of study.

There is another topic in this paper that is related to personal incomes,
but in a juridical sense, namely, company incomes. It was recently reported
by Okuyama and Takayasu, who analyzed high quality Japanese company data 
for the
year 1997, that the income distribution of companies accurately follows 
Zipf's law, that is, a cumulative distribution with
exponent $-1$ \cite{ott}. 
It is anticipated that this law holds rather universally
for different years, for different countries, and even for each job category,
with some exceptions. The second aim of this paper is to show a
complementary result relating to the company income distributions, that is,
the distribution of debts of bankrupt companies, which may correspond to
the distribution of negative incomes. Although the data available for this
purpose is limited we can apparently demonstrate that the debts distribution
also follows the same statistics, Zipf's law.

In the following section we give the details of our personal income
distribution analysis. We briefly describe the result of the debt
distribution in section 3, and the final section is devoted to
discussions of the implications and significance of our results.

\section{The income and income-tax distributions}
The data we have for income and income tax
are in many senses complementary.
Both data relate to Japan, and the income data is for the fiscal year 
1997 and 1998, while the income-tax data is only for 1998.
For the fiscal year 1998,
the income data contains all 6,224,254 workers
who filed tax returns, but it is a coarsely tabulated data, 
while the income-tax data lists the income tax of individuals 
who paid tax of ten million yen or more in the same year.
In this section we study the distribution of each data set and
then combine them to obtain an over-all picture of
income distribution in high income range.

The tabulated data for the income distribution of individuals 
in Japan is publicly available on the web pages of the Japanese
Tax Administration \cite{webincome}. Since 
the relevant pages are in Japanese only, we quote the data for
the fiscal year 1998 in Table \ref{tbl:income},
where and hereafter we denote the income by $s$ in units of million yen.
It should be noted that this data is for individuals who filed a
tax return, and therefore does not include all individuals with income.
Under the Japanese tax system, 
if a worker has only one source of income (salary) and
the income is less than 20,000,000 yen, that person does not have to
file a tax return (the tax deducted from the monthly salary is
adjusted at the end of the year by the employer and that
becomes the final amount of tax paid).
In some circumstances,  
even if the income is below this amount, a person may have to
file a tax return if some other conditions are met.
Consequently, a large number of individuals in fact submit returns. 
Therefore for $s< 20$ this data represents only a 
portion of the actual number of individuals, and we expect that
the deviation between the number of people in this table
and the actual number of people in that income range
becomes larger in lower income ranges.
Rigorously speaking, 
it is safe to trust only the entries for income $s \ge 20$
(the last three columns) in this data
as a faithful representation of the number of individuals in 
that income range. 

\begin{table}[ht]
\begin{center}
\begin{tabular}{|c|r|}\hline
Income ($s$-million yen)& Number\\ \hline
$s \le 0.7$ & 14,496\\ \hline
$0.7 < s \le 1$ & 73,352\\ \hline
$1 < s \le 1.5$ & 359,157\\ \hline
$1.5 < s \le 2$ & 541,739\\ \hline
$2 < s \le 2.5$ & 565,835\\ \hline
$2.5 < s \le 3$ & 584,989\\ \hline
$3 < s \le 4$ & 954,901\\ \hline
$4 < s \le 5$ & 687,057\\ \hline
$5 < s \le 6$ & 497,438\\ \hline
$6 < s \le 7$ & 375,485\\ \hline
$7 < s \le 8$ & 288,141\\ \hline
$8 < s \le 10$ & 379,716\\ \hline
$10 < s \le 12$ & 229,205\\ \hline
$12 < s \le 15$ & 215,712\\ \hline
$15 < s \le 20$ & 190,524\\ \hline
$20 < s \le 30$ & 140,533\\ \hline
$30 < s \le 50$ & 82,514\\ \hline
$50 < s $& 43,455\\ \hline
\end{tabular}
\caption{The tabulated data of the income of 6,224,254 
individuals in Japan in fiscal year 1998.}
\label{tbl:income}
\end{center}
\end{table}

\begin{figure}[ht]
\begin{center}
\includegraphics[width=13cm]{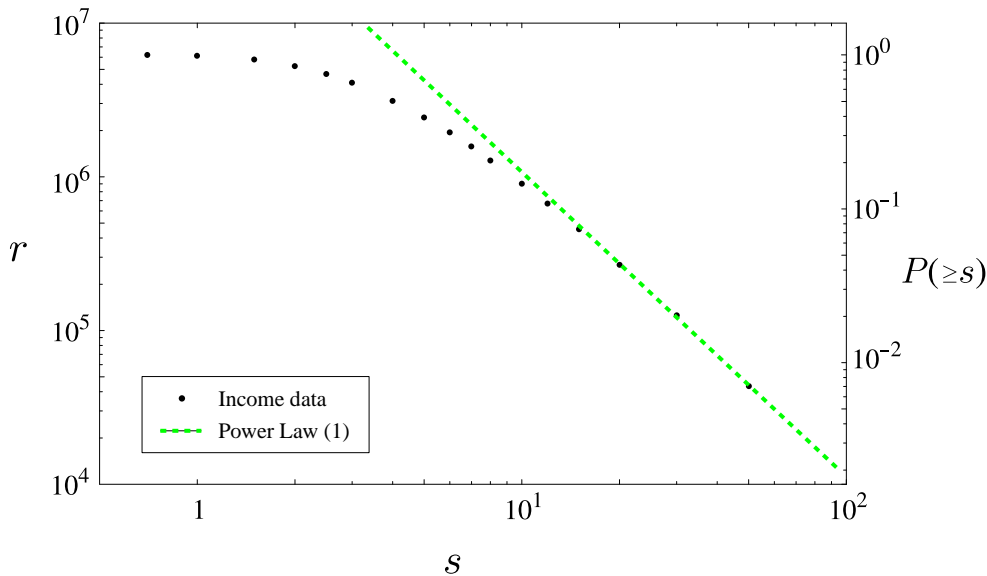}
\caption{The rank-size plot of the income distribution of Table \ref{tbl:income}
(dots) and the linear approximation (broken line) given by 
Eq.(\ref{eqn:trustme}).}
\label{fig:income}
\end{center}
\end{figure}

In Fig.~\ref{fig:income} we give the rank-size plot, which is the 
log-log plot of the rank ($r$) as a function of income ($s$).
The vertical axis may be converted to the accumulated 
probability $P(\ge s)$, the probability that 
a given individual has income equal to or greater than $s$.
This is done by simple rescaling $P(\ge s)=r(s)/r(0)$ where $r(0)=6,224,254$,
and therefore it merely induces a translation of the vertical axis 
in the log plot and does not change the shape of the distribution.
The ticks for the accumulated probability $P(\ge s)$ are
placed on the right-hand side of the plot.
It is apparent that this plot tends to a linear function
in the high income range, say $s>10^1$.
As we have noted above, this distribution should be equal to
that of all the workers only for $s\ge 20$.
Therefore we have calculated the best-fit linear function
from the last three data points and have found that it is given
by the following;
\begin{equation}
r \propto s^{-1.98}.
\label{eqn:trustme}
\end{equation}
The broken line in Fig.\ref{fig:income} represents this function.
The distribution $P(\ge s)$ fits
the power law (\ref{eqn:trustme}) very well in the high income range
and it gradually deviates from (\ref{eqn:trustme}) as the income
becomes lower. Since this deviation starts to grow as the
income $s$ becomes lower than 20, we may assume that
the power law (\ref{eqn:trustme}) applies to a much wider range
of incomes. 

\begin{figure}[ht]
\begin{center}
\includegraphics[width=11cm]{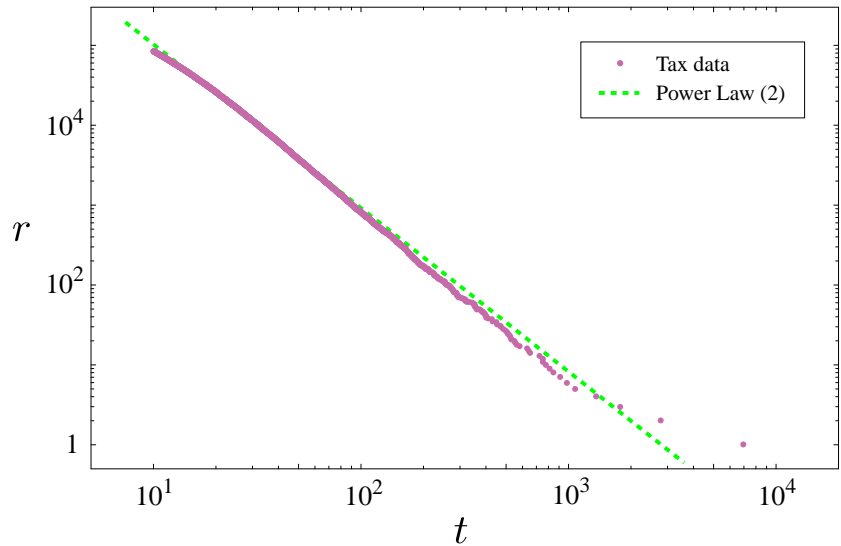}
\caption{The rank-size plot of the income-tax listing (dots) and the
function (\ref{eqn:tax}) obtained by the best-fit method (broken line).}
\label{fig:tax}
\end{center}
\end{figure}

Another set of publicly available data is a list of
all taxpayers who paid income-tax of ten million yen or more in 
1998.  Hereafter we denote the tax by $t$ in million yen.
There are 84,515 such individuals listed for fiscal year 1998.
The rank-size plot of this data is given in Fig.\ref{fig:tax}.
This distribution is very close to being 
linear in this log-log plot. The best-fit result is
given by the following;
\begin{equation}
r \propto t^{-2.05}.
\label{eqn:tax}
\end{equation}
This function is given by the dotted line in Fig.\ref{fig:tax}.

The fact that the income-tax distribution satisfies Eq.(\ref{eqn:tax}),
which is very close to the income distribution Eq.(\ref{eqn:trustme}),
implies that income-tax is roughly proportional to the income.
This proportionality constant can be fixed by 
assuming that the income-tax is a 
monotonically increasing function
of the income, which we expect to be true in an averaged sense.
By this assumption a given person has the same rank in both
the income listing and the income-tax listing.
Therefore the relation between 
income and income-tax can be found in the overlapping
region of those two data.
The data of incomes, however, is given only in approximate form 
by Table \ref{tbl:income}, while
the listing of the income-tax covers the bottom one and a half columns.
Therefore, the comparison between the income and the income-tax
is possible only for the rank $r=43,455$ whose income was $s=50$.
From our listing of the income-tax we find that the 43,455th ranked
person has paid income tax of $t=15.13$.
Therefore we find that 
\begin{equation}
t = 0.30 \, s.
\label{eqn:st}
\end{equation}
In the fiscal year 1998, the tax ($t$) was obtained from the income ($s$)
in three steps: (1) various deductions, such as the basic (default) deduction,
deductions for the spouse and the dependents, and insurance deductions,
were made to calculate the taxable income ($s'$),
(2) the basic tax ($t'$) is calculated according to the
formula $t'=0.5 s' - 6.03$ in this income range,
and (3) further deductions, such as the deduction for donation to political
parties, deduction for housing loans and further special deductions 
are made to arrive at the actual tax $t$.
Although it is difficult to assess the validity of the result (\ref{eqn:st}),
it may be understood as an averaged result of these steps, especially since
the proportionality constant is meaningfully less than the constant 0.5 of
the tax formula.

\begin{figure}[ht]
\begin{center}
\includegraphics[width=11cm]{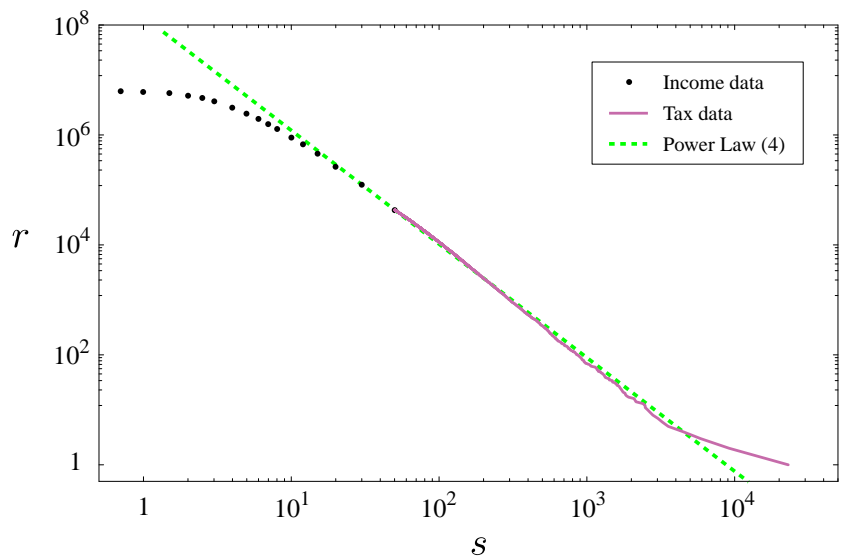}
\caption{The rank-size plot of the income. The data in Table \ref{tbl:income}
is shown by dots as in Fig.\ref{fig:income}, 
whereas the data points induced from the income-tax listing by 
Eq.(\ref{eqn:st}) are connected by the solid line-segments.
The broken line shows the function (\ref{eqn:allzipf}) obtained by
the best-fit method.}
\label{fig:allzipf}
\end{center}
\end{figure}

Using the result (\ref{eqn:st}), we translate the income-tax data to
the income data. This way, the shortcoming of the original
income data, that it lacks the information in the high income
range, can be overcome. The resulting rank-size plot is given in 
Fig.\ref{fig:allzipf}.

Again, the power-law behavior is apparent in Fig.\ref{fig:allzipf}.
Caution is needed in applying the best-fit method;
The data points of the tax-induced income 
becomes very dense at lower $s$, therefore they would dominate the
result if all data points are treated with equal weight
in a simple best-fit method.
The data of original income table would certainly be neglected in such
an analysis.
For this reason, we have calculated the best fit to the power law 
from each of the two columns in Table \ref{tbl:income}\
and each of the five sections $[50\times 2^n, 50\times2^{n+1}] (n=1 \cdots 5)$
of the tax-induced income data and have taken the
average of the seven linear functions (in the log-log scale).
The result is the following;
\begin{equation}
r \propto s^{-2.06}.
\label{eqn:allzipf}
\end{equation}
The broken line in Fig.\ref{fig:allzipf} denotes this function.
We find that the power-law (\ref{eqn:allzipf}) fits the data
over three magnitude of income, $10<S<10^4$.

As mentioned earlier, the deviation of the income data points
from the power law in the range $s < 20$ may be explained as 
the result of the exclusion of the majority of the 
workers of single salary source.  
On the other hand, there is a possibility 
that the distribution in the lower ranges
follow the log-normal distribution as was suggested by
Montroll and Shlesinger \cite{lognormal}.
Therefore it is highly desirable 
to investigate these lower ranges of income to obtain the 
overall profile of the income distribution.
One such study toward this direction is currently in progress
by one of the authors (W.S.) and will be published
in near future \cite{souma}.

\section{Debts of bankrupt companies}

\begin{figure}[ht]
\begin{center}
\includegraphics[width=11cm]{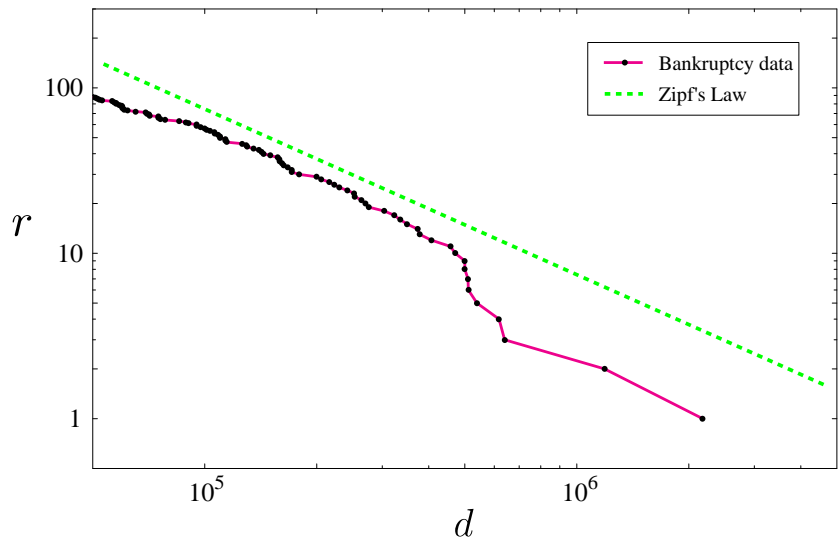}
\caption{The rank-size plot of the accumulated probability
$P(\ge d)$ of the debt ($d$) from 1997 to March, 2000.
The broken line is the Zipf law (\ref{eqn:zipf})
shown for visual reference.}
\label{fig:tosan}
\end{center}
\end{figure}

The data we have gathered covers some 100 big bankruptcies in Japan from
1997 to the end of March, 2000.
Unfortunately the exact criteria for the choice of the ``big"
cases is not known.
The resulting rank-size plot for the accumulated probability
$P(\ge d)$ is given in Fig.\ref{fig:tosan}, where we have
denoted the debt in units of million yen by $d$.
The data actually contains much smaller cases, but from the fact that
the number of data starts to decrease for $d<10^5$ we
assume that the data starts to loose reliability in this range.
The broken line represents the Zipf law,
\begin{equation}
r \propto d^{-1},
\label{eqn:zipf}
\end{equation}
which were found for the company income previously \cite{ott}.
This data is in reasonable agreement with this power-law.

\section{Discussion}
As described in section 2 we have discovered a new type of income
distribution for individuals, the power law distribution with an exponent close
to $-2.0$. If we assume that all the pioneering results together with our
result, then the only consistent interpretation may be as follows: Income
distribution generally follows a power law for large incomes but its
exponent varies from country to country and year to year. 
As noted, Pareto and Gini give
different exponents using data from different years. However, there is a
possibility that such old results may based on imperfect data, and we may
still expect a kind of universal behavior for personal income distributions.
In our analysis the observed scale range satisfying the power law in Fig.3 is
about three decades in the income rank range ($10$ to $10^4$), therefore, the
estimation of the exponent is quite accurate. We need comparably high
precision data for other countries or years for further discussion about such
universality.

There has been no established theory for the income distributions. A
classical theory was proposed by Gibrat in 1932 \cite{refGib}. 
He assumed the time
evolution of each person's income to be approximated by a multiplicative
stochastic process considering that a random process proportional to the
amount of the present income can approximate the increase and decrease of
income. By this assumption the resulting distribution of income follows a
log-normal distribution which is consistent with the result of Montroll and
Shlesinger for salaried people. However, this theory apparently fails to
explain the more interesting part of the distribution, the power law
tails.

Explaining the existence of power law tails with the level of
simplicity in Gibrat's theory can easily be done as follows by introducing a
simple modification to his model. Let us introduce an additive random term
together with the multiplicative term, such as
\begin{equation}
I(t+1) = a(t)I(t) +  R(t).
\end{equation}
If we assume that $a(t)$ and $R(t)$ are independent random variables then it
is mathematically shown that the distribution of $I(t)$ follows a 
general power law
even in a case where the fluctuation of $a(t)$ is
correlated \cite{refsato}. 
In the simplest case when both $a(t)$ and $R(t)$ are white
noises, the power law exponent, $\beta$, of the distribution of $I(t)$ 
is simply
given by the following formula \cite{reftaka},
\begin{equation}
\langle a^\beta \rangle=1.
\end{equation}
Based on this relation the power law distribution with an anticipated
exponent can be obtained by only tuning the statistics of growth rate, 
$a(t)$.

There can be many different approaches other than this simple growth
model. One straightforward model can be given by relating the company's
incomes to personal incomes by assuming that the high income individuals are
mostly company owners. If we can assume that company owner's income
is proportional to the square root of the company's income, then the
established empirical Zipf's law for the company's income reduces the
squared power law for individuals.

Nonlinear interaction models of incomes may also be important such as
the model introduced by Levy and Solomon \cite{ls}. An obvious difficulty in
such modeling is the mathematical form of interaction among people working
in real economics. For example economical interaction such as buying and
selling generally occur in a discrete manner both in time and price,
therefore, it is even problematic whether we can introduce an analytical
function as an interaction term.

Related to the issue of the income distribution and the dynamics behind it
is the distribution of wealth of individuals.
One such data set, from the Forbes 400, is analyzed in Ref.\cite{ls2},
and is shown to have the Pareto exponent close to $-1.36$.
We have recently carried out a similar analysis with 1996--1999 data of
approximately 400 of the richest people of U.S.A. and 100 of the 
richest people in Japan
and found that the results are consistent with this exponent.

As there is no steady model of personal incomes it is important to
approach from many different aspects. Personal income might be related to a
kind of popularity such as the citation numbers in scientific publications
or access number distribution for Internet homepages. Citation frequency
distribution has been examined for publications in the field of physics and
a power law tail with exponent about $-3$ is reported \cite{redner}. As for
homepage popularity ranking it is known that we have another Zipf's law for
this quantity, namely the power law exponent is about $-1$ \cite{refhomepage}.

Our other finding concerning the distribution of bankrupt companies is
considered to be suggestive for the understanding of money flow statistics
among companies. The amount of debt of a company is roughly given by
accumulated negative incomes that exceeded the whole asset of the company.
Therefore, combining our result with the known Zipf's law for positive
incomes we may conjecture that the distributions of money flows among
companies both positive and negative generally tend to follow Zipf's
law.

For the purpose of construction of a numerical model of company
income statistics, we can summarize the known conditions to be satisfied as
follows:
\begin{enumerate}
\item The amount of annual summation of outgoing money from each company is of
the same order of its assets, also the summation of incoming money is of the
same order of the assets \cite{ott}.
\item By definition a company's income is given by the difference of these
moneys, that is, incoming money minus outgoing money. The averaged incomes is
nearly proportional to a fractional power of asset with exponent estimated
to be about 0.85 \cite{ott}.
\item The distribution of income follows Zipf's law for both positive and
negative regions.
\item The variance of growth fluctuation of assets is proportional to a
fractional power of assets with exponent about $-0.15$ \cite{refstan}.
\item The distribution of assets seems to follow a log-normal law 
\cite{refamaral},
but we need better quality real data to assess this empirical law.
\end{enumerate}

Although no model has successfully satisfied all of these conditions,
there are two promising approaches towards this goal. One is a company interaction
model based on a zero-sum game \cite{reftaka} 
introduced to explain the asset's
variance law 4. In this model companies play a kind of stochastic game in
which money flow from losers to winners according to a certain rule. In this
model the condition 4 and 5 are satisfied in a rough sense, however, condition
3 is not fulfilled. The condition 1 depends on the definition of the annual
year, which can be tuned, while condition 2 has not been checked.

The other approach is a territory occupation model that is based on the
assumption that a company's income may be proportional to the share of
customers \cite{refbook}. Companies under certain stochastic rules in 
this model
share the customers in a new field of a market. 
From this approach it is easy to
satisfy Zipf's law for positive incomes. However, other conditions are
out of the range at present.

\vspace{3mm}
\noindent {\bf\large Acknowledgments}\\
The authors would like to thank Mr.~Nakano of the Japanese
Tax Administration for conversation on the properties of the data
on the web pages.
Numerical computation in this work was in part supported by
the computing facility at the Yukawa Institute for 
Theoretical Physics.
The authors would also like to thank Dr.~John Constable 
(Magdalene College, U.K.) for careful reading of the manuscript.

\end{document}